\documentclass[comment, prl,twocolumn,nofootinbib, preprintnumbers, superscriptaddress]{revtex4}

\usepackage{amsmath,amssymb,bm,slashed,braket}
\usepackage{graphicx}
\usepackage{epstopdf}
\usepackage{float}
\usepackage{xcolor}
\usepackage{multirow}
\usepackage{dsfont}
\usepackage[colorlinks=true,
            linkcolor=blue,
            urlcolor=blue,
            citecolor=green,          
            bookmarks=true,
            bookmarksnumbered=true,
            breaklinks=true,
            pdfpagemode=Fullscreen,
            pdfstartview=FitBH]{hyperref}
\usepackage[normalem]{ulem}
\allowdisplaybreaks[4]
\usepackage[capitalise]{cleveref}

\newcommand{\calO}{ {\cal O} }
\newcommand{\calS}{ {\cal S} }
\newcommand{\pve}{{\pmb{v}_{\rm el}^\perp}}

\begin{document}

\title{Revisiting general dark matter-bound-electron interactions}

\author{Jin-Han Liang}
\email{jinhanliang@m.scnu.edu.cn}
\author{Yi Liao}
\email{liaoy@m.scnu.edu.cn}
\author{Xiao-Dong Ma}
\email{maxid@scnu.edu.cn}
\author{Hao-Lin Wang}
\email{whaolin@m.scnu.edu.cn}
\affiliation{Key Laboratory of Atomic and Subatomic Structure and Quantum Control (MOE), 
Guangdong Basic Research Center of Excellence for Structure and Fundamental Interactions of Matter, 
Institute of Quantum Matter, South China Normal University, Guangzhou 510006, China}
\affiliation{Guangdong-Hong Kong Joint Laboratory of Quantum Matter, 
Guangdong Provincial Key Laboratory of Nuclear Science, 
Southern Nuclear Science Computing Center, South China Normal University, Guangzhou 510006, China}

\begin{abstract}
In this Letter we revisit general dark matter (DM)-bound-electron interactions studied previously in the influential work [R. Catena {\it et al.,} Atomic responses to general dark matter-electron interactions, \href{https://journals.aps.org/prresearch/pdf/10.1103/PhysRevResearch.2.033195}{Phys.\,Rev.\,Res.\,\textbf{2},\,033195,\,(2020)}]. 
For the most general DM-electron nonrelativistic or relativistic interactions for DM with spin up to 1, 
we find the average ionization matrix element squared can be organized into three terms, each of which is a product of a DM response function ($a_{0,1,2}$) and a linear combination ($\widetilde W_{0,1,2}$) of the four atomic response functions ($W_{1,2,3,4}$) given in that work,
\vspace{-0.2cm}
\begin{equation*}
\widetilde W_0 = W_1, \,
\widetilde W_1 = |\bm{v}_0^\perp|^2 W_1 - 2 {m_e\, \bm{q}\cdot \bm{v}_0^\perp \over \bm{q}^2} W_2 + W_3,\,
\widetilde W_2 = { (\bm{q}\cdot \bm{v}_0^\perp)^2 \over \bm{q}^2} W_1 
- 2 {m_e\, \bm{q}\cdot \bm{v}_0^\perp \over \bm{q}^2} W_2 + {m_e^2 \over \bm{q}^2}W_4. 
\vspace{-0.2cm}
\end{equation*}%
Furthermore, we find a crucial minus sign was missed for the calculation of $W_2$ in that work, which has significant phenomenological consequences when explaining experimental bounds on specific DM scenarios. 
Due to the corrected sign, there can be significant cancellations between the $W_2$ and $W_{3,4}$ terms, so that $\widetilde W_{1,2}$ are dominated by the usual response function $W_1$ in some cases. 
Many DM scenarios involving DM or electron axial-vector current can yield $W_2$ and thus are potentially affected by the sign. As an example, we show that the recent XENON1T constraint on the fermionic DM anapole moment is weakened by a factor of 2 or so. 
We also present a complete list of NR operators for spin-1 DM and compute their contributions to the DM response functions.

\end{abstract}

\maketitle 

%%%%%%%%%%%%
{\it Introduction.} 
The nature of dark matter (DM) is one of 
the biggest puzzles in particle astrophysics. Particle DM is a 
well-motivated scenario that fits very well within the standard 
cosmological model, and in the meanwhile has a detectable possibility in terrestrial experiments. 
The mass scale for potential DM candidates spans a very wide range of energy scales, from the superlight axion or dark photon candidate \cite{Duffy:2009ig,Fabbrichesi:2020wbt,Agrawal:2021dbo}, to the keV sterile neutrino \cite{Drewes:2016upu, Boyarsky:2018tvu}, to relatively more massive particles that are generally termed weakly interacting massive particles (WIMPs) \cite{Arcadi:2017kky,Roszkowski:2017nbc}. Among them, the WIMPs are the most prominent ones in consideration of their origin in particle physics models, the power of fitting the observational data, and the promising discovery
potential in direct detection experiments. 

The search for WIMP DM begins with the nuclear recoil effect. 
The null results from large xenon and argon detectors 
(such as PandaX \cite{PandaX-II:2017hlx}, XENON \cite{XENON:2023cxc}, LUX-ZEPLIN \cite{LZ:2022lsv}, and DarkSide \cite{DarkSide:2018bpj}) in the past two decades have put stringent bounds on the DM-nucleus cross section with DM mass above 
$\calO (10\,\rm GeV)$ \cite{Roszkowski:2017nbc,Schumann:2019eaa}. 
Due to the kinematic restriction of DM-nucleus (or DM-nucleon) elastic scattering as well as detection threshold limitation, 
the DM-nucleon interaction remains less constrained for DM mass below ${\cal O}(10\,\rm GeV)$ from nuclear recoil signals. 
This limitation could be circumvented by exploiting inelastic processes like the bremsstrahlung process \cite{Kouvaris:2016afs} and the Migdal effect \cite{Migdal:1941,Ibe:2017yqa}, or novel low-threshold detectors in condensed matter systems (see \cite{Kahn:2021ttr,Mitridate:2022tnv} and references cited therein). 
In contrast to nuclear recoil experiments searching for the DM-nucleon interaction, 
the DM-electron interaction offers a more powerful alternative probe of low mass DM particles 
below $\calO(10\,\rm GeV)$ through the electron recoil signal
\cite{LZ:2023poo,XENON:2019gfn,XENON:2021qze,XENON:2022ltv,PandaX:2022xqx,DarkSide:2018ppu,PandaX-II:2021nsg}.
This is due to the significantly smaller mass of the target electron, allowing it to gain more easily the recoil energy from a light DM particle. 
For instance, the single-electron search conducted by XENON1T has the capability of exploring DM as light as approximately 5 MeV \cite{XENON:2021qze}.

Since the search for light DM through DM-electron scattering has become more and more important than ever before, a thorough understanding of the response functions of the electron in a target material is urgent. The dual-phase detectors,
with liquid xenon or argon as the target material, are the most promising DM direct detection experiments to explore the parameter space of WIMPs in a wide mass region, and can effectively distinguish between the signals induced by the DM-electron and DM-nucleus scattering. Theoretically, the DM-electron scattering in a liquid xenon target was first calculated in \cite{Kopp:2009et,Essig:2011nj,Essig:2012yx}, in which some simplified DM models were studied, with the atomic effect of bound electrons being encoded in the ionization form factor \cite{Essig:2011nj} or the $K$ factor \cite{Roberts:2016xfw}. The relativistic correction and many-body effect of the ionization form factor were further considered in \cite{Roberts:2015lga,Roberts:2016xfw,Pandey:2018esq},
which could be sizable for a large transfer momentum. Recently, a general study on the atomic response functions in the liquid xenon and argon targets was carried out in \cite{Catena:2019gfa} in the framework of nonrelativistic (NR) effective field theory (EFT), in which three new atomic response functions were introduced in addition to the usual ionization form factor. 
These new atomic response functions appear for general DM-electron interactions and are particularly relevant for some special DM scenarios. 
They have been used in subsequent theoretical \cite{Catena:2022fnk} and experimental \cite{XENON:2021qze} papers. 
In this work, we rederive these response functions, and find that a crucial minus sign was missed for the second response function $W_2$ defined in \cite{Catena:2019gfa}. The minus sign is physically important when explaining experimental bounds in terms of specific DM scenarios. 
For the purpose of illustration and the interest in its own right, we study the constraints on the DM anapole operator which were considered previously in \cite{Catena:2022fnk,XENON:2021qze}. We find that the constraints are weakened by a factor of 2 or so upon correcting the sign of $W_2$. 
Along the line we formulate the squared matrix element of the DM-atom scattering from the most general NR EFT operators for DM up to spin 1, and find that it can be compactly organized into three terms  [see \cref{eq:M12sq2}], each of which is a product of a DM response function and a linear combination of atomic response functions.

%%%%%%%%%%%%
\vspace{0.5cm}{\it Formalism for DM-atom scattering.}
For the DM-electron scattering in atomic targets, we label the initial state of the bound electron by the usual atomic quantum numbers $(n,\ell,m)\equiv|1\rangle$ 
and the final state of the ionized electron by $(k^\prime,\ell^\prime,m^\prime)\equiv|2\rangle$ with $k'$ being its momentum. 
Denoting the initial (final) three-momentum of DM by $\bm{p}\,(\bm{p}')$, the transition amplitude for $|\bm{p}, 1\rangle\to | \bm{p}', 2\rangle $ can be written as \cite{Catena:2019gfa} 
\begin{align}
\label{eq:amplitude1}
{\cal M}_{1\to 2} =\int \frac{{\rm d}^3 \bm{k} }{(2 \pi)^3} \tilde\psi_{2}^*(\bm{k}+\bm{q}) 
 {\cal M}(\bm{q}, \pve) \tilde\psi_1(\bm{k}), 
\end{align}
where $\tilde \psi_1\equiv \tilde \psi_{n\ell m}$ and 
$\tilde \psi_2 \equiv \tilde \psi_{k'\ell' m'}$ are the initial and final electron wave functions in momentum space respectively, 
and are related to the wave functions in configuration space by the Fourier transformation, 
$\tilde \psi(\bm{k}) = \int {\rm d}^3 \bm{r}\, e^{-i \bm{k}\cdot\bm{r}} \psi(\bm{r})$. 
${\cal M}(\bm{q}, \pve)$, ${\cal M}$ for short, is the matrix element for free electron-DM scattering
that is only a function of the momentum transfer 
$\bm{q}\equiv \bm{p}-\bm{p}'$ 
and the DM-electron transverse velocity 
$\pve=\bm{v}-{\bm{q}/(2\mu_{xe})}-{\bm{k}/ m_e}$
in the nonrelativistic framework. Here $\bm{v}$ is the incoming DM velocity
in the lab frame, $\bm{k}$ is the momentum of the initial electron of mass $m_e$, 
and $\mu_{xe}$ the reduced DM-electron mass, respectively. 

For the most general DM-electron interactions at leading order in the NR EFT framework (see the Supplementary Material)
which are at most linear order in $\pve$ or $\bm{k}$, 
the free amplitude is decomposed as \cite{Catena:2019gfa} 
${\cal M}={\cal M}_{\pmb{k}=0}+\pmb{k}\cdot(\nabla_{\bm{k}} {\cal M})_{\pmb{k}=0}$. 
This results in two atomic form factors $f_{1 \to 2}(\bm{q})$ and $\pmb{f}_{1 \to 2}(\bm{q})$ in \cref{eq:amplitude1}:
\begin{subequations}
\begin{align}
f_{1 \to 2}(\bm{q})&= \int \frac{{\rm d}^3 \bm{k}}{(2 \pi)^3} \tilde\psi_2^*(\bm{k}+\bm{q}) \tilde \psi_1(\bm{k}), 
\\
\pmb{f}_{1 \to 2}(\bm{q})&= \int \frac{{\rm d}^3 \bm{k}}{(2 \pi)^3} \tilde\psi_2^*(\bm{k}+\bm{q}){\bm{k} \over m_e} \tilde\psi_1(\bm{k}).    
\end{align}
\end{subequations}
The first one is the usual ionization form factor, while the second was first recognized in \cite{Catena:2019gfa}. 
We find that a minus sign was missed in the vector form factor $\bm{f}_{1 \to 2}(\bm{q})$ of \cite{Catena:2019gfa} 
when transforming from momentum to configuration space, 
\begin{subequations}
\begin{align}
f_{1 \to 2}(\bm{q}) & = 
\int {\rm d}^3 \bm{r}\, \psi^*_2 (\bm{r}) e^{i\bm{q}\cdot\bm{r}} \psi_1(\bm{r}),
\\
\pmb{f}_{1 \to 2}(\bm{q}) &= 
\int {\rm d}^3 \bm{r} \,\psi^*_2 (\bm{r})e^{i\bm{q}\cdot\bm{r}}
{- i\pmb{\nabla} \over m_e} \psi_1(\bm{r}).
\end{align}    
\end{subequations}
This will cause a considerable impact on the matrix element squared as spelled out below. 

The decomposition of ${\cal M}$ in terms of $\bm{k}$ leads the authors in \cite{Catena:2019gfa} to identify three additional atomic response functions $W_{2,3,4}$ beyond the usual $W_1$. From their definitions (collected in Supplementary Material), one sees that the minus sign in $\pmb{f}_{1 \to 2}(\bm{q})$ is brought into the
function $W_2$. 
We realize that it is more advantageous to expand ${\cal M}$ in $\bm{v}_{\rm el}^\perp$ as this makes power counting manifestly consistent with that of NR operators:
\begin{align}
\label{eq:decom2}
{\cal M}(\bm{q}, \pve) 
& =  {\cal M}_{\tt S}
+  \bm{v}_{\rm el}^\perp\cdot
{\pmb{\cal M}}_{\tt V},
\end{align}
where ${\cal M}_{\tt S} = {\cal M}(\bm{q}, 0)$, and 
${\pmb{\cal M}}_{\tt V} = \nabla_{\bm{v}_{\rm el}^\perp} {\cal M}(\bm{q}, \pve)$ is actually independent of $\bm{v}_{\rm el}^\perp$ for our consideration of NR operators up to its linear order. This expansion also helps to appreciate the consequence of the sign issue. Then,
\begin{align}
 {\cal M}_{1\to 2} = f_{\tt S}(\bm{q}) {\cal M}_{\tt S}
 +  \bm{f}_{\tt V}(\bm{q})\cdot {\pmb{\cal M}}_{\tt V},
\end{align}
where $f_{\tt S}= f_{1\to2}(\bm{q})$ and
\begin{align}
 \bm{f}_{\tt V} =\int {{\rm d}^3 \bm{k} \over (2\pi)^3 }\tilde\psi_2^*(\bm{k}+\bm{q})\bm{v}_{\rm el}^\perp \tilde\psi_1(\bm{k}).    
\end{align}
With the above decomposition,
we find the spin-averaged atomic matrix element squared takes the form
\begin{align}
 \overline{|{\cal M}_{1\to2}|^2} 
 \label{eq:M12sq}
& = a_0 |f_{\tt S}|^2
+ a_1 |\bm{f}_{\tt V}|^2
+ {a_2 \over x_e}  \left|{\bm{q}\over m_e} \cdot \bm{f}_{\tt V} \right|^2
\\
& + i\, a_3 {\bm{q} \over m_e}\cdot(\bm{f}_{\tt V}\times \bm{f}_{\tt V}^*)
+ 2\, {\rm Im}\left[a_4 f_{\tt S} \bm{f}_{\tt V}^* \cdot {\bm{q}\over m_e} \right],
\nonumber
\end{align}
where $ x_e\equiv \bm{q}^2/m_e^2$. 
$a_{0,1,2,3,4}$ are DM response functions depending only on the Wilson coefficients (WCs) of NR operators and $x_e$. 
The explicit forms of $a_{0,1,2}$ are provided in the Supplementary Material
for the scalar, fermion, and vector DM cases, respectively.
The other two terms $a_{3,4}$ have no contribution to the usual DM-atom scattering after summing over the atomic magnetic quantum numbers $(m,m')$, 
and will be presented in \cite{Liang:2024ecw}. 

With $ \overline{|{\cal M}_{1\to2}|^2}$ in \cref{eq:M12sq}, 
the total ionization rate 
for the initial atomic orbital ($n,\ell$),
${\cal R}_{\rm ion }^{n \ell}$, 
can be obtained by summing over
the initial magnetic quantum number $m$ 
and integrate/sum over all the
allowed final electron states ($k^\prime,\ell^\prime,m^\prime$).
Neglecting the effect arising from the nonspherically symmetric nature 
of the DM velocity distribution in the lab frame,
the differential rate can be generally written as \cite{Essig:2015cda,Catena:2019gfa}
{\small
\begin{align}
\label{eq:Rion}
\frac{{\rm d} {\cal R}_{{\rm ion}}^{n \ell}}{{\rm d} \ln E_e}=\frac{n_{\rm dm}}{128\pi m_{\rm dm}^2 m_e^2}
\int {\rm d}q \, q
\int \frac{{\rm d}^3 \bm{v} }{v} f_{\rm dm}(\pmb{v})  \overline{\left|{\cal M}_{{\rm ion}}^{n \ell}\right|^2},
\end{align}
}%
where $n_{\rm dm}$ is the local DM number density with velocity distribution $f_{\rm dm}(\bm{v})$ and $m_{\rm dm}$ is the DM mass.
The integration over $v$ is bounded below by $v_{\rm min}$ due to the kinematics restriction.
Using the decomposition in \cref{eq:decom2} and the squared matrix element in \cref{eq:M12sq}, 
our main result in this Letter is the recognition of the averaged ionization matrix element squared, 
\begin{align}
\label{eq:M12sq2}
\overline{\left|{\cal M}_{\rm{ion}}^{n \ell}\right|^2}=
a_0 \widetilde W_0 
+ a_1 \widetilde W_1 
+ a_2 \widetilde W_2. 
\end{align}
Denoting the $\bm{k}$-independent part of $\bm{v}_{\rm el}^\perp$ as $\bm{v}_0^\perp\equiv \bm{v}-{\bm{q}/(2\mu_{xe})}$, we thus have 
$ \bm{f}_{\tt V} \equiv \bm{v}_0^\perp f_{1\to2}(\bm{q}) - \bm{f}_{1\to2}(\bm{q})$, which leads to the coefficients 
$\widetilde W_{0,1,2}$ as linear combinations of atomic response functions $W_{1,2,3,4}$ in \cite{Catena:2019gfa},\footnote{Notice that both $\widetilde W_i$ and $W_i$ depend on the initial atomic quantum numbers $(n,\ell)$, the final electron momentum $k'$, and the momentum transfer $|\bm{q}|$. We do not show them explicitly for brevity.}
\begin{subequations}
\label{eq:Wtilde}
\begin{align}
\widetilde W_0 &= W_1,
\\
\widetilde W_1 &= |\bm{v}_0^\perp|^2 W_1 -2 {y_e \over x_e} W_2 + W_3, 
\\
\widetilde W_2&= {y_e^2 \over x_e} W_1  -2  {y_e\over x_e} W_2 + {1\over x_e}W_4.
\end{align}
\end{subequations}
where $y_e \equiv  \bm{q}\cdot \bm{v}_0^\perp/ m_e$.

We find the decomposition in \cref{eq:decom2} is more amenable to obtain the square of the ionization matrix element with $a_0$ and $a_{1,2}$ capturing different NR contributions: while $a_0$ captures only the velocity independent operators, $a_{1,2}$ are sensitive only to the velocity dependent ones.
Furthermore, the recognition of the three new combinations of atomic response functions is useful to understand which NR operators contribute dominantly to the scattering. 
In \cref{eq:M12sq2}, $\widetilde W_0$ (or $W_1$) is the usual atomic $K$ factor in the literature \cite{Roberts:2016xfw,Roberts:2019chv}. 
Its associated DM response function $a_0$ incorporates the most common DM scenarios through the WCs $c_1$ and $c_4$ of the NR operators $\calO_1$ and $\calO_4$, which are the well-known spin-independent and spin-dependent NR interactions extensively studied in the literature. 
$\widetilde W_{1,2}$ are mainly related to the NR operators involving transverse velocity $\pve$, 
which can be generated by DM-electron interactions involving electron or DM axial-vector current such as the anapole moment of fermionic DM. 
Thus, a sign mistake in $W_2$ will result in a wrong interpretation of experimental data for such types of interactions.

Let us compare our formalism with the one in \cite{Catena:2019gfa}. 
Our $\widetilde{W}_{0,1,2}$ 
contain three parameters: $x_e$, $y_e$, and $|\bm{v}_0^\perp|^2$. The parameter $x_e$ is independent of DM. The parameter $y_e= \Delta E/ m_e - x_e/2$, 
which is 
also independent of DM in spite of its definition involving $\bm{v}_0^\perp$. 
Here $\Delta E$ is the energy level gap between the initial and final electron orbitals. 
The only dependence on DM is 
in the parameter $|\bm{v}_0^\perp|^2$ for $\widetilde W_1$, but is very weak. As 
shown in \cite{Liang:2024ecw}, 
\begin{align}
|\bm{v}_0^\perp|^2
= v^2 + {q^2 \over 4 m_e^2 }\left(1 - 
{m_e^2\over m_x^2} \right) - {\Delta E\over m_e}\left(1 + {m_e \over m_x } \right),
\end{align}
where $v\sim 10^{-3}$ is the typical DM velocity. For the interested range of DM mass in liquid xenon/argon experiments ($m_x \gtrsim 5\,{\rm MeV} \sim 10\, m_e$), the relative correction due to DM mass is less than 1\,\% due to the dominance of the second term, and thus is safely negligible. 
In contrast, in the formalism of \cite{Catena:2019gfa} which factorizes the rate into a product of DM response functions (called $R_i^{nl}$ in that work) and atomic functions $W_{1,2,3,4}$, the former ones depend on the atomic orbitals manifestly (see their Eqs.\,(39) and (40) and Appendix C) with contributions up to 40\,\% for some terms \cite{Liang:2024ecw}.   
In summary, while our linear combinations $\widetilde{W}_{0,1,2}$ depend very weakly on DM through its mass and velocity, the DM response functions in \cite{Catena:2019gfa} depend significantly on the energy gap between the initial and final atomic orbitals.

\begin{figure}[t]
\centering
\includegraphics[width=1\linewidth]{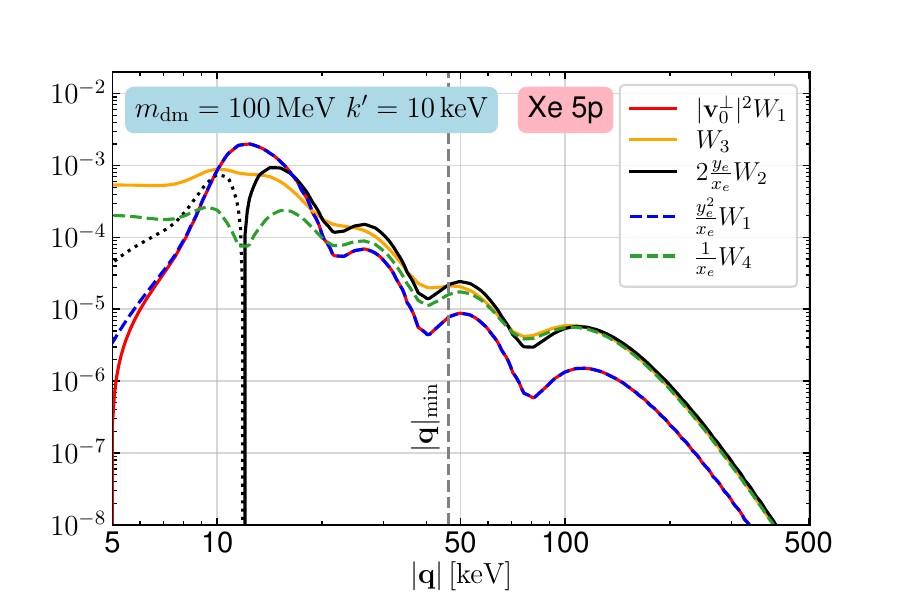}
\caption{Individual terms in $\widetilde W_{1,2}$ are shown as a function of $|\pmb{q}|$.
For better layout we show $- 2(y_e/x_e)W_2$ (black dotted curve) for $|\bm{q}|\lesssim 11.5\,\rm keV$
and $+2(y_e/x_e)W_2$ (black solid) for $|\bm{q}|\gtrsim 11.5\,\rm keV$. 
The vertical dashed gray line with $|\bm{q}|_{\rm min}$ indicates the minimum $|\bm{q}|$ to ionize a 5p electron with an outgoing momentum of 10 keV.
}
\label{fig:Wi}
\end{figure} 

To assess the numerical importance of the sign, we show in \cref{fig:Wi} the individual terms of $\widetilde W_{1,2}$ as a function of the momentum transfer $|\bm{q}|$ for a xenon target in the orbital 5p. We choose a typical light DM mass $m_{\rm dm}=500\,\rm MeV$ 
and fix the ionized electron momentum to be $k'=10\,\rm keV$. 
From \cref{fig:Wi}, it is clear that the $W_2$ term is positive for $|\bm{q}| \gtrsim 11.5\,\rm keV$ which covers the experimentally relevant region $|\bm{q}|\ge|\bm{q}|_{\rm min}$, 
and approaches $W_3$ and $W_4/x_e$ as $|\bm{q}|$ increases. 
With the minus sign accompanying $W_2$ in $\widetilde W_{1,2}$, this results in a significant cancellation between the $W_2$ and $W_{3,4}$ terms in $\widetilde W_{1,2}$, 
leaving the $W_1$ term as the potentially dominant contribution across a wide region of $|\bm{q}|$. 
We find that the cancellation behavior is 
observed across different atomic orbitals, DM masses, ionized electron momenta, and targets (e.g., argon), though its extent varies, leaving behind only about 0.1\,\% to 10\,\% of the contribution from $W_2$ and $W_3$.
It occurs even in the presence of a light mediator. This is because the experimentally relevant region $|\bm{q}|\ge|\bm{q}|_{\rm min}$ renders the low $|\bm{q}|$ region less relevant to the event rate.

\begin{figure}[t]
\centering
\includegraphics[width=1\linewidth]{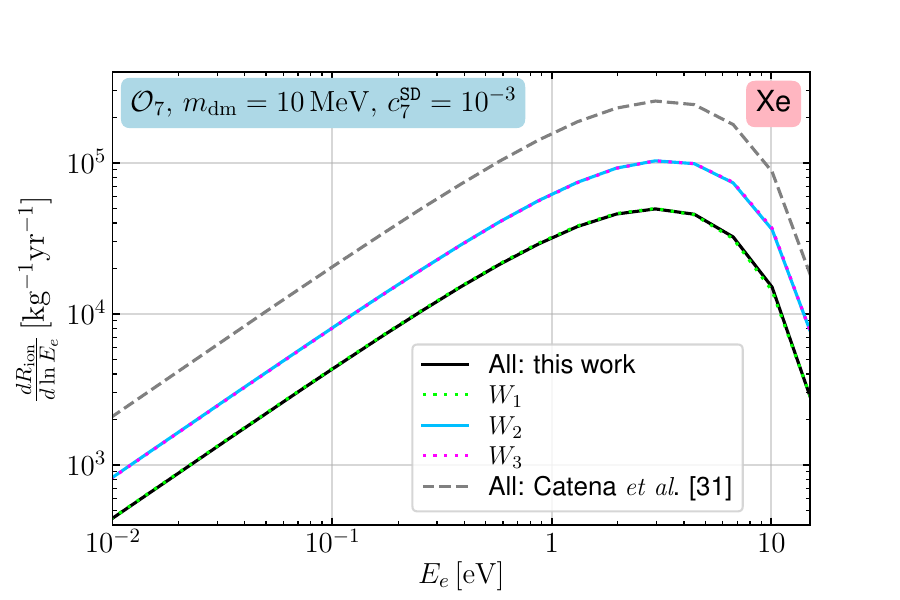}
\caption{The ionization spectrum from the NR interaction ${\cal O}_7$ with a short distance origin at the xenon target from different response functions. 
The DM can be either a scalar, fermion, or vector particle. Note that the solid black and dotted green curves almost overlap.
}
\label{fig:dRdE}
\end{figure}

\begin{figure}[t]
\centering
\includegraphics[width=1\linewidth]{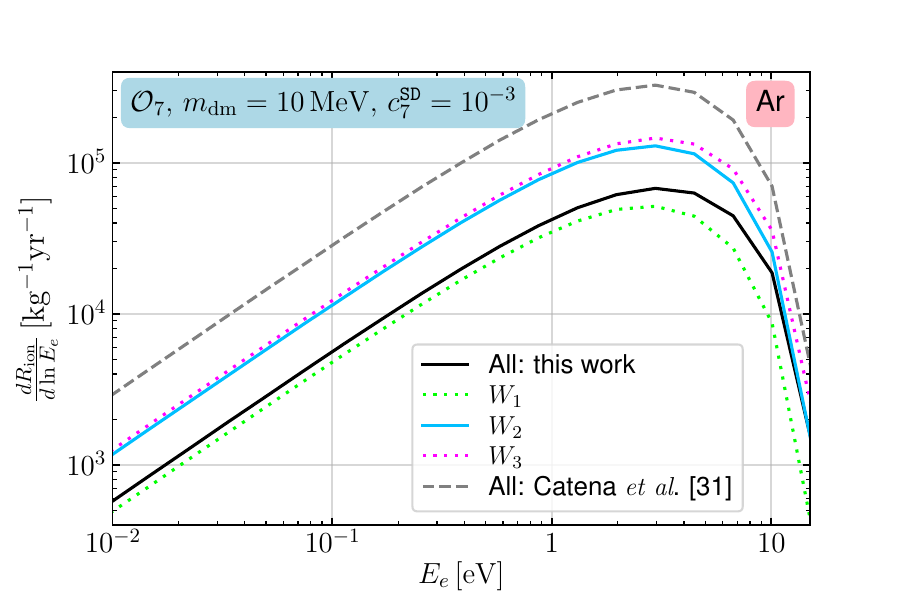}
\caption{Same as \cref{fig:dRdE} except for an argon target.}
\label{fig:dRdEAr}
\end{figure}

Next we examine the influence on the differential rate due to the NR operator $\calO_{7}$, 
which is a typical operator employed in \cite{Catena:2019gfa} to show the importance of the three new response functions ($W_2,W_3,W_4$).
To facilitate comparison with \cite{Catena:2019gfa}, we consider the xenon and argon targets and focus on the contact (or short-distance) interaction with 
$c_{7}^{\tt SD}=10^{-3}$ and $m_{\rm dm}=10\,\rm MeV$. In \cref{fig:dRdE}, we show for the xenon target the individual contribution to the differential rate from $W_1$, $W_2$, and $W_3$ [including all coefficients but not the minus sign in front of $W_2$ in \cref{eq:Wtilde} for better layout] by the dotted green, solid blue, and dotted magenta curves. From \cref{fig:dRdE}, it can be seen that the contributions from $W_2$ and $W_3$ are of a similar magnitude and much larger than that from $W_1$,
so that they almost exactly cancel out, 
leading to a total contribution (solid black curve) that is completely dominated by $W_1$.   
For comparison, we also show the total rate reported in \cite{Catena:2019gfa} by the dashed black curve.\footnote{There is about a global factor of 3/4 difference than the one given in \cite{Catena:2019gfa}, which has been included in our plot.}
It can be seen that the sign difference can cause a factor of 5 difference in the differential rate.
A similar plot for the argon target is provided in \cref{fig:dRdEAr}. 
In that case, the cancellation between $W_2$ and $W_3$ is less complete, leading to a slightly higher total contribution than that from $W_1$ alone. 
Note that the spectrum is the same for NR operator $\calO_8$ with fermion DM, 
which will be used in the following analysis. 

%%%%%%%%%%%%
\vspace{0.5cm}
{\it Constraints on DM anapole.}
In this part, we use the DM anapole operator as an example to illustrate the importance of the sign for $W_2$. 
The anapole interaction Lagrangian for a fermionic DM $\chi$ is
\begin{align}
\mathcal{L}_{\rm anapole} = \frac{g}{2\Lambda^2} \bar{\chi} \gamma^\mu \gamma^5 \chi
\partial^\nu F_{\mu\nu}, 
\end{align}
where $F_{\mu\nu}$ is the electromagnetic field strength tensor, 
$g$ is a dimensionless coupling, and $\Lambda$ is a heavy particle scale. 
We consider $\chi$ as a Majorana particle for direct comparison with the results in the literature \cite{Catena:2019gfa,XENON:2021qze}. 
The anapole moment is the unique electromagnetic property of a Majorana fermion since all others (such as charge radius and electromagnetic dipole moments) do not exist. 
In the NR limit, the DM anapole operator matches to two NR operators, $\mathcal{O}_8$ and $\mathcal{O}_9$,
with the WCs,
\begin{align}
c_8 = -c _9 =  8\, e\, m_e m_\chi {g\over \Lambda^2}. 
\end{align}

Since the sign of $W_2$ has a significant impact on the contribution from $\calO_8$, as illustrated in \cref{fig:dRdE},
it is expected that it also plays a crucial role in determining the constraints on the DM anapole operator. 
The previous constraints were established using data from various experiments, 
including XENON10, DarkSide-50, the XENON1T S2-only search (calculated in \cite{Catena:2019gfa}), 
and the XENON1T single electron (SE) search conducted by the XENON1T collaboration \cite{XENON:2021qze}. 
All of these constraints were obtained with a wrong sign associated with $W_2$. 
In the following, we analyze these constraints from xenon target experiments. 
In addition, we calculate the constraint using the latest S2-only data from the PandaX-4T experiment \cite{PandaX:2022xqx}.

To compare with the S2-only data observed in the experiments, we utilize the constant-W model \cite{Essig:2017kqs, Essig:2012yx} to convert the ${\rm d} {\cal R}_{{\rm ion}}^{n \ell} / {\rm d} E_e$ spectrum into that of the number of ionized electrons, ${\rm d} {\cal R}_{{\rm ion}}^{n \ell} / {\rm d} n_e$. 
For each ionized electron, we assume that the number of photoelectrons (PEs) induced by it follows a Gaussian distribution with a mean value $g_2$ 
and a width value $\sigma_{\rm S2}$ as in \cite{Essig:2017kqs, Essig:2012yx}. Specifically, for XENON10 (XENON1T), we take $g_2 =27\,(33)$ and $\sigma_{\rm S2}=6.7\,(7)$ \cite{Essig:2017kqs,Essig:2019xkx}. By convoluting with the Gaussian distribution \cite{Essig:2012yx, Essig:2017kqs, XENON:2019gfn}, we obtain the spectrum of S2 signals (the number of PEs), 
${\rm d} {\cal R}_{{\rm ion}}^{n \ell} / {\rm d PE} $. 
For the constraints from the XENON10 and XENON1T S2-only searches, we employ the same efficiency factors and statistical methods as in \cite{Catena:2019gfa}. 
With a wrong sign of $W_2$ we were able to reproduce the constraints in \cite{Catena:2019gfa} to check our numerical analysis. 

XENON1T-SE:  
Since the XENON1T collaboration did not provide all the necessary information in their paper \cite{XENON:2021qze}, 
it is challenging to reproduce its constraint with high consistency. 
Therefore, we adopt a rescaling method described below to update the constraint with the corrected sign of $W_2$. 
The signal region in the XENON1T SE search is selected as $\rm S2 \in [14,150]\,\rm PE$, corresponding to signals with 1--5 ionized electrons \cite{XENON:2021qze}. For a specific DM mass, we find that the spectrum ${\rm d} {\cal R}_{{\rm ion}}^{n \ell} / {\rm d} n_e$ with the correct-sign $W_2$ differs from that with the wrong-sign $W_2$ by nearly a global factor. For instance, for $m_\chi =1\,\rm GeV$, this global factor is approximately 3 for the anapole operator.
Hence, for each DM mass, we calculate $\sum_{n_e=1}^5{\rm d} {\cal R}_{{\rm ion}}^{n \ell} / {\rm d} n_e$ with both the wrong-sign and correct-sign $W_2$. 
Then, we deliver our new constraint by rescaling the constraint of the XENON1T collaboration.

PandaX-4T: 
With the exposure of 0.55 ton$\cdot$yr, 103 events (including 95.8 background events estimated) were observed in the signal region of S2 $\in [60,200]\,\rm PE$ \cite{PandaX:2022xqx}.
According to the Poisson distribution, a 90\,\%  C.L. requires the number of signal events to be less than 21.5.
We employ the total efficiency (red solid line) given in Fig.\,1 of \cite{PandaX:2022xqx}. 
Since the parameters $g_2$ and $\sigma_{\rm S2}$ of the PandaX-4T S2-only search were not given explicitly in the paper, we read from its Fig.\,3 $g_2=17.7$ and assume $\sigma_{\rm S2}=\sqrt{g_2}$.

\begin{figure}
\centering
\includegraphics[width=1\linewidth]{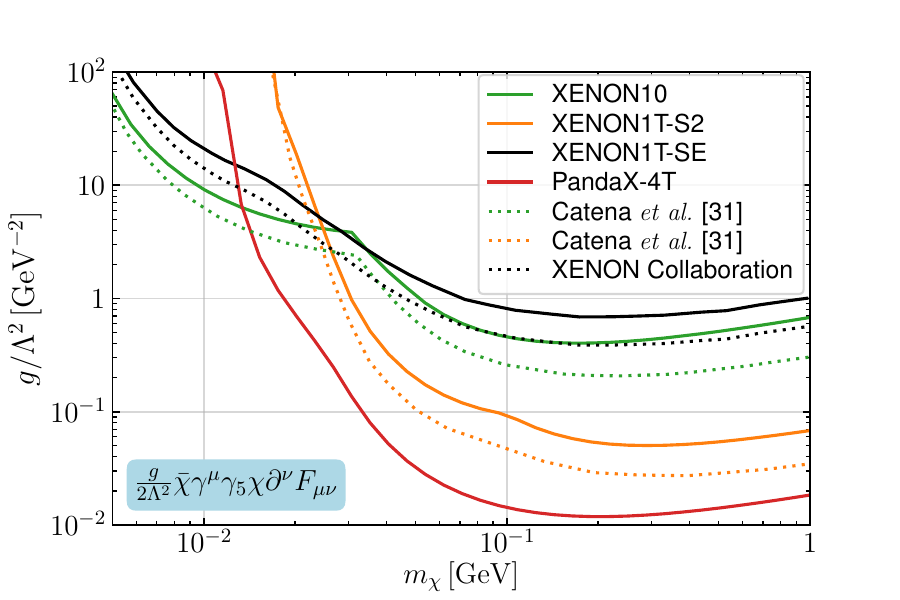}
\caption{Constraints on DM anapole moment from DM direct detection experiments,
including XENON10, XENON1T S2-only search, XENON1T-SE search,
and PandaX-4T. The solid lines show our results with a correct-sign $W_2$, 
while the dashed lines show the previous constraints \cite{Catena:2019gfa,XENON:2021qze} with a wrong-sign $W_2$.
}
\label{fig:anapole}
\end{figure}

To ensure consistent comparison with previous calculations based on a wrong-sign $W_2$, 
we adopt the same DM parameters as \cite{Catena:2019gfa}: $\rho_\chi = 0.4 \,\rm GeV/cm^3$ for the local DM density, 
$v_0 =220\,\rm km/s$ for the Sun's circular velocity, $v_{\rm esc} = 544 \,\rm km/s$ for the galactic escape velocity, 
and $v_\oplus = 244 \,\rm km/s$ for the speed of Earth in the galactic rest frame, respectively. 
Our results on XENON10, XENON1T S2, XENON1T SE, and the latest PandaX-4T data are shown in \cref{fig:anapole} with solid curves. 
Correcting the $W_2$ sign, the constraints are weakened by a factor of $\sim 2$, depending on the DM mass. 
With a larger DM mass, the discrepancy tends to be larger. The constraint derived from the latest PandaX-4T data imposes the most stringent limit for DM mass 
$ m_\chi \gtrsim 20\,\rm MeV $, reaching $g/\Lambda^2 \lesssim 0.01\,\rm GeV^{-2}$ for $m_\chi \gtrsim 0.1\,\rm GeV$. 
In the low-mass region with $m_\chi \lesssim 20\,\rm MeV$, the stronger limit is set by the previous XENON10 and XENON1T-SE experiments, which have a low energy threshold down to one single ionized electron.

%%%%%%%%%%%%
\vspace{0.5cm}{\it Conclusion.}                                     
In this Letter, we studied the general DM-electron interactions and found, for the most general leading-order nonrelativistic dark matter-electron interactions, 
that those interactions naturally organize themselves into the differential rate in the form of three DM response functions $a_{0,1,2}$. 
Each of $a_{0,1,2}$ is multiplied by a coefficient $\widetilde W_{0,1,2}$ which is a linear combination of the four atomic response functions $W_{1,2,3,4}$ defined in \cite{Catena:2019gfa}. Most importantly, we found a crucial sign mistake associated with $W_2$. 
Our results show that this leads to important phenomenological implications. 
As far as general NR operators are concerned, it will influence the interpretation of experimental data on the NR operators entering $a_{1,2}$, i.e., $\calO_{7,8,12,13,14}$ and $\calO_{17,21,22,23,25,26}$ for DM up to spin 1. 
Furthermore, at the more fundamental level of relativistic field theory, several interesting DM scenarios are particularly affected by the sign of $W_2$.
Especially, the interactions with electron or DM axial-vector or tensor currents can induce contributions related to $W_2$, 
and thus would be affected by the sign problem.
Lastly, our results present a complete list of NR operators for spin-one DM and their contribution to the DM response functions $a_{0,1,2}$.
The calculation details in this Letter, 
together with the systematic constraints on other nonrelativistic and relativistic interactions from the latest experiments, 
will be provided in the long paper \cite{Liang:2024ecw}. 

%%%%%%%%%%%%%%%%%%%%%%%%%%%%%% 
\section*{Acknowledgements}
%%%%%%%%%%%%%%%%%%%%%%%%%%%%%%
This work was supported in part by the Guangdong Major Project of Basic
and Applied Basic Research No.\,2020B0301030008, and by the Grants 
No.\,NSFC-12035008, 
No.\,NSFC-12247151, 
No.\,NSFC-12305110,
and No.\,NSFC-12347121. 

\bibliography{refs.bib}{}
\bibliographystyle{utphys}

\newpage\null\newpage
%%%%%%%%%%%%%%%%%%%%%%%%%%%%%%
\onecolumngrid
%%%%%%
\begin{center}
 {\large Supplementary Material}
\end{center}
%%%%%%%

%%
{\pmb{Nonrelativistic operators for DM-electron scattering}---}% 
The NR DM-electron interactions can be built by using the identity and spin operators. For the electron (and fermion or vector DM), we denote the identity and spin operators by $\mathds{1}_e$ and $\bm{S}_e$ ($\mathds{1}_x$ and $\bm{S}_x$), 
$x=\phi, \chi, X$ for the scalar, fermion, vector DM cases respectively. For the
vector DM, besides $\mathds{1}_x$ and $\bm{S}_x$, a rank-two traceless spin tensor operator $\pmb{\tilde\calS}_x$ is needed:
\begin{align}
\tilde{\calS}_x^{ij}=\frac{1}{2}\left(S_x^i S_x^j+i \leftrightarrow j\right)-\frac{2}{3}\delta^{ij}.
\end{align}
Together with the momentum transfer $\pmb{q}$ and DM-electron transverse velocity $\pve$, one can construct the relevant NR operators. 
In \cref{tab:NRope}, we summarize the most general NR operators for DM-electron scattering at leading order. 
The NR operator basis is given for operators that are at most quadratic in $\bm{q}$ and linear in $\pve$ for a DM particle up to spin one. 
The operators from $\calO_1$ to $\calO_{15}$ were given previously in Refs.\,\cite{Catena:2019gfa,Catena:2022fnk}, and the operators from $\calO_{17}$ to $\calO_{26}$ are specific to vector DM-electron scattering which can be generated by the relativistic interactions in \cite{He:2022ljo,Liang:2024ecw}. 
We note in passing that the basis of operators for DM-quark interactions can be obtained by changing the electron label to the flavor of quarks.

\begin{table}[h]
\centering
\begin{tabular}{| l | l | l |}
\hline
$\calO_1 =\mathds{1}_x \mathds{1}_e$
& $\calO_{10} =  \mathds{1}_x \, \frac{i \bm{q}}{m_e}\cdot \bm{S}_e$
& $\calO_{19} =  \frac{\bm{q}}{m_e} \cdot \pmb{\tilde\calS}_x \cdot  \frac{\bm{q}}{m_e} \mathds{1}_e$
\\
$\calO_3 = \mathds{1}_x \Big(\frac{i \bm{q}}{m_e}\times \pve \Big) \cdot  \bm{S}_e$        
&  $\calO_{11} =  \bm{S}_x \cdot \frac{i\bm{q}}{m_e} \mathds{1}_e$
& $\calO_{20} =  - \frac{\bm{q}}{m_e}  \cdot \pmb{\tilde\calS}_x \cdot \Big(\frac{\bm{q}}{m_e} \times  \bm{S}_e\Big)$
\\
$\calO_4 = \bm{S}_x \cdot \bm{S}_e$         
&  $\calO_{12} = - \bm{S}_x \cdot (\pve\times \bm{S}_e)$
& $\calO_{21} = \pve\cdot\pmb{\tilde\calS}_x\cdot\bm{S}_e$
\\
$\calO_5 = \bm{S}_x \cdot \Big(\frac{i\bm{q}}{m_e}\times \pve\Big) \mathds{1}_e$         
&  $\calO_{13} = (\bm{S}_x \cdot \pve) \Big(\frac{i\bm{q}}{m_e}\cdot \bm{S}_e  \Big)$
& $\calO_{22} = \Big(\frac{i \bm{q}}{m_e}\times\pve\Big)\cdot\pmb{\tilde\calS}_x\cdot\bm{S}_e + \pve\cdot\pmb{\tilde\calS}_x\cdot\Big(\frac{i \bm{q}}{m_e}\times\bm{S}_e\Big)$
\\
$\calO_6 =  \Big(\bm{S}\cdot \frac{\bm{q}}{m_e}\Big) \Big( \frac{\bm{q}}{m_e} \cdot \bm{S}_e \Big)$         
&  $\calO_{14} = (\bm{S}_x \cdot \frac{i\bm{q}}{m_e} ) (\pve \cdot \bm{S}_e)$
& $\calO_{23} = - \frac{i \bm{q}}{m_e}\cdot\pmb{\tilde\calS}_x\cdot(\pve \times \bm{S}_e)$
\\
$\calO_7 = \mathds{1}_x \, \pve \cdot \bm{S}_e$         
&  $\calO_{15} = \bm{S}_x \cdot \frac{\bm{q}}{m_e} \Big[ \frac{\bm{q}}{m_e}  \cdot ( \pve \times \bm{S}_e)\Big]$
& $\calO_{24} = {\bm{q} \over m_e} \cdot \pmb{\tilde\calS}_x \cdot  \Big({ \bm{q} \over m_e} \times\pve\Big)$
\\
$\calO_8 = \bm{S}_x \cdot \pve \, \mathds{1}_e$         
&  $\calO_{17} = \frac{i \bm{q}}{m_e} \cdot \pmb{\tilde\calS}_x \cdot \pve \,\mathds{1}_e$
& $\calO_{25}  = \Big({\bm{q}\over m_e} \cdot \pmb{\tilde\calS}_x\cdot\pve \Big) 
\Big({\bm{q}\over m_e} \cdot\bm{S}_e \Big)$
\\
$\calO_9 = -\bm{S}_x \cdot\Big(\frac{i\bm{q}}{m_e}\times \bm{S}_e \Big)$         
&  $\calO_{18} = \frac{i \bm{q}}{m_e} \cdot \pmb{\tilde\calS}_x \cdot  \bm{S}_e$
&$\calO_{26} = \Big({\bm{q}\over m_e} \cdot \pmb{\tilde\calS}_x \cdot {\bm{q}\over m_e} \Big) (\pve\cdot\bm{S}_e)$
\\
\hline
\end{tabular}
\caption{The leading-order nonrelativistic DM-electron interactions. 
Here $\pmb{q}=\pmb{p}-\pmb{p}'$ with $\pmb{p}$ ($\pmb{p}'$) being the momentum of initial (final) DM state. 
$\calO_2$ and $\calO_{16}$ being quadratic in $\pve$ are not shown.}
\label{tab:NRope}
\end{table}

{\pmb{Dark matter response functions}---}% 
The calculation details for DM response functions are given in \cite{Liang:2024ecw}. For the DM up to spin one, the DM response functions are
\begin{itemize}
\item Scalar DM:
\begin{align}
\label{eq:scalarDMRF}
a_0  & = c_1^2+ {1\over 4} c_{10}^2 {\color{cyan} x_e},  
& a_1  & = {1\over 4} c_7^2 + {1\over 4} c_3^2 {\color{cyan}x_e}, 
& a_2  & = - {1\over 4} c_3^2 {\color{cyan}x_e}. 
\end{align}   
\item Fermion DM: 
\begin{subequations}
\label{eq:FermionDMRF}
\begin{align}
a_0 & = c_1^2+\frac{3 }{16} c_4^2 +
\left( {1\over8}c_9^2+{1\over4} c_{10}^2+{1\over4} c_{11}^2 + {1\over8} c_4 c_6\right) {\color{cyan} x_e }
+{1\over16} c_6^2 {\color{cyan} x_e^2 }, 
\\
a_1 & =  {1\over4} c_7^2+{1\over4} c_8^2+{1\over8}c_{12}^2 
+\left( {1\over4} c_3^2+ {1\over4} c_5^2+ {1\over16} c_{13}^2+ {1\over16} c_{14}^2
-{1\over8} c_{12} c_{15}\right) {\color{cyan}  x_e }
+\frac{1}{16} c_{15}^2 {\color{cyan}  x_e^2} ,
\\
a_2  & = - \left( {1\over4} c_3^2+ {1\over4}  c_5^2- {1\over8}  c_{13} c_{14}- {1\over8} c_{12} c_{15}\right){\color{cyan}x_e}  - \frac{1}{16} c_{15}^2 {\color{cyan}x_e^2} .
\end{align}
\end{subequations}
\item Vector DM: 
\begin{subequations}
\label{eq:VectorDMRF}
\begin{align}
a_0  & =  c_1^2+\frac{1}{2} c_4^2
+\left({1\over3} c_9^2+{1\over4}c_{10}^2+{2\over3}c_{11}^2 +{5\over36} c_{18}^2 
+{1\over3} c_4 c_6\right) {\color{cyan} x_e}
+ \left({1\over6} c_6^2+{2\over9} c_{19}^2+{1\over12} c_{20}^2\right) {\color{cyan}  x_e^2 },
\\%
 a_1 &   = {1\over4} c_7^2+{2\over3} c_8^2+{1\over3} c_{12}^2 +{5\over36} c_{21}^2 
+ \left({1\over4} c_3^2+{2\over3} c_5^2+{1\over6} c_{13}^2+{1\over6} c_{14}^2+{1\over6} c_{17}^2 +{3\over8} c_{22}^2 +{7\over72} c_{23}^2 
-{1\over3} c_{12} c_{15} \right.
\nonumber
\\
& \left.  +{1\over12} c_{22} c_{23}  +{1\over12} c_{21} c_{25}-{1\over18} c_{21} c_{26}\right) {\color{cyan} x_e} 
+\left({1\over6} c_{15}^2+{1\over6} c_{24}^2+{1\over24} c_{25}^2+ {1\over18} c_{26}^2\right) {\color{cyan} x_e^2},  
\\%
a_2  & = 
- \left( {1\over 4} c_3^2+ {2\over3} c_5^2-{1\over18}  c_{17}^2+ {7\over 24}c_{22}^2+ {1\over 72}  c_{23}^2-{1\over3} c_{13} c_{14}-{1\over3} c_{12} c_{15}  - {1\over36}  c_{21} c_{25} -{1\over 6} c_{21} c_{26} +{1\over4}  c_{22} c_{23} \right) {\color{cyan} x_e} 
\nonumber
\\
&
-  \left( {1\over 6}   c_{15}^2+{1\over 6}  c_{24}^2- {1\over 72}  c_{25}^2 
- {1\over 9} c_{25} c_{26} \right) {\color{cyan} x_e^2}. 
\end{align}
\end{subequations}
\end{itemize}
In the above, we highlight the dependence on $x_e=\bm{q}^2/m_e^2$ in cyan 
that helps to identify the NR operators with leading order contributions. The above results are for the case of real WCs $c_i$ that correspond to Hermitian operators for the elastic DM scenario. For the complex WCs, the results are obtained by changing $c_ic_j\to {\rm Re}(c_i^*c_j)$ in each term. The case with complex WCs could appear in the inelastic DM scenario, all of which will be provided in \cite{Liang:2024ecw}.  

{\pmb{Atomic response functions}---}% 
The definition for the atomic response functions $W_{1,2,3,4}$ is as follows \cite{Catena:2019gfa}, 
\begin{subequations}
\begin{align}
W_1 & = V {4 k'^3 \over {(2 \pi)^3}} \sum_{m=-\ell}^{\ell} \sum_{\ell'=0}^\infty \sum_{m'=-\ell'}^{\ell'} |f_{1\to 2}(\bm{q})|^2,
\\
W_2 & = V {4 k'^3 \over {(2 \pi)^3}} \sum_{m=-\ell}^{\ell} \sum_{\ell'=0}^\infty \sum_{m'=-\ell'}^{\ell'}  {\bm{q} \over m_e} \cdot f_{1\to 2}(\bm{q}) \bm{f}_{1\to2}^*(\bm{q}),
\\
W_3 & = V {4 k'^3 \over {(2 \pi)^3}} \sum_{m=-\ell}^{\ell} \sum_{\ell'=0}^\infty \sum_{m'=-\ell'}^{\ell'} |\bm{f}_{1 \to 2}(\bm{q})|^2,
\\
W_4 & = V {4 k'^3 \over {(2 \pi)^3}} \sum_{m=-\ell}^{\ell} \sum_{\ell'=0}^\infty \sum_{m'=-\ell'}^{\ell'} |{\bm{q} \over {m_e}} \cdot \bm{f}_{1\to 2}(\bm{q})|^2.
\end{align}
\end{subequations}
Here the summations have been performed over the magnetic quantum number ($m$) for the initial electron state and the angular and magnetic quantum numbers ($\ell'$ and $m'$) for the final electron state,
and $V$ is the normalization volume. 
Notice the definition of $W_{1,2,3,4}$ are independent of specific DM properties. Although the three new combinations of atomic response functions ($\widetilde W_{0,1,2}$) are related to the DM mass through $|\bm{v}_0^\perp|^2$, we emphasize that they are useful for understanding which NR operators give
dominant contributions, as discussed in the paper.

\end{document}